\documentclass[11pt]{article}
\usepackage[dvips]{graphicx,color}
\usepackage{amsmath}
\usepackage{amsfonts}
\usepackage[latin1]{inputenc}

\begin{document}



\title{PROPERTIES OF MAGNETIZED QUARK-HYBRID STARS}

\author{ M. ORSARIA$^{1, 2, 3}$, IGNACIO F. RANEA-SANDOVAL$^{1, 2}$,\\
H. VUCETICH$^{1}$ and F. WEBER$^{3}$\\
$^1$Gravitation, Astrophysics and Cosmology Group \\
Facultad de Ciencias Astron{\'o}micas y Geofísicas,\\ Paseo del Bosque S/N (1900),\\
Universidad Nacional de La Plata UNLP, La Plata, Argentina\\
$^2$ CONICET, Rivadavia 1917, 1033 Buenos Aires, Argentina\\
$^3$ Department of Physics, San Diego State University,\\ 
5500 Campanile Drive, San Diego, CA 92182, USA}
\date{}
\maketitle



\begin{abstract}
  The structure of a magnetized quark-hybrid stars (QHS) is modeled
  using a standard relativistic mean-field equation of state (EoS) for
  the description of hadronic matter. For quark matter we consider a
  bag model EoS which is modified perturbatively to account for the
  presence of a uniform magnetic field.  The mass-radius (M-R)
  relationship, gravitational redshift and rotational Kepler periods
  of such stars are compared with those of standard neutron stars
  (NS).
\end{abstract}

\section{Introduction}

It is known that compact objects such as NS or hybrid stars
possess enormous magnetic fields. Anomalous X-ray pulsars (AXPs) and
soft $\gamma$-repeaters (SGRs), may contain NS with magnetic fields
greater than $10^{15}$~G at the NS surface (magnetars). Some authors
\cite{Cheng,Ouyed} claim that magnetized QHS or magnetized
quark stars (QS) might be the real sources of such objects.

In a recent paper \cite{Orsaria} we modeled QS as high-density quark
bags with magnetic fields of $B \sim 4-6 \times 10^{17}$~G. Although
such magnetic fields are typical for magnetars, it was shown that the
magnetic field is still low enough so that it can be treated
perturbatively ({\it i.e.}, $B \gg \mu^2$, with $\mu$ being the
baryon chemical potential).

In this paper we analyze the structure of a QHS,
consisting of a strange quark matter (SQM) core enveloped in a thin
hadron matter crust. We neglect the narrow gap that exits between the
SQM core and the crust, and use an EoS for such stars which describes
confined hadronic matter in terms of nucleons and hyperons (HV of Ref.\
\cite{Fridolin0}) and deconfined quark matter in terms of a
relativistic gas of up, down and strange quarks, as described by the
modified MIT bag model of Ref. \cite{Orsaria}.

\section{High density magnetized SQM EoS}

For the quark matter in the core of a QHS we consider massless quarks,
which implies that the electrons are not present and the quark
chemical potentials are, as a consequence of chemical equilibrium, all
equal, $\mu_u=\mu_d=\mu_s \equiv \mu$. Considering the limit of weak
magnetic fields, $\mu^2 \gg B$, after some analytic approximations the
EoS of magnetized SQM within the framework of the MIT Bag model
becomes \cite{Orsaria}
\begin{equation}
  \label{rho} \rho \simeq 3 P + 4 \mathrm{B}_{\mathrm{eff}}-\frac
  {B^2}{3{\pi}^{2}}\,\left(2-\mathrm{Log}\frac{B}{2^{1/3}\,3\,\mu^2}\right),
\end{equation}
with $\mathrm{B}_{\mathrm{eff}}\,=\, \frac{B^2}{8 \pi}\,+\,B_{\rm
  bag}\,$, where $B_{\rm bag}$ denotes the bag constant.  Bag values
in the range of $57~ {\rm MeV~ fm}^{-3} < B_{\rm bag} < 90~ {\rm MeV~
  fm}^{-3}$ correspond to SQM which is absolutely stable with respect
to nuclear matter \cite{Fridolin0,Farhi84:a}, even when the magnetic
field $B \neq 0$ \cite{Orsaria}. For QHS, which are made of
meta-stable SQM, we consider $B_{\rm bag} = {\rm 120~ MeV~ fm}^{-3}$.
For such a value of the bag constant the threshold for the magnetic
field is $B_{\rm max}=6.4 \times 10^{17}$~G.

\section{Redshift and Keppler period }

Several EoS for NS, QHS and QS have been proposed but none of them is
conclusive \cite{Haensel}. Each EoS produces a different mass-radius
(M-R) relationship which can be contrasted with the available
observational data in order to test its range of validity and/or set
bounds on some parameters. The gravitational surface redshift is of
particular interest for distinguishing between QHS and NS because it
is an observable quantity.  The surface redshift, $z$, depends on the
mass $M$ and radius $R$ of the star and is given by
\begin{equation}
  z = (1 - 2GM/Rc^2)^{-1/2} - 1.
\end{equation}
We also consider the effect of rotation and calculate the maximum
possible rotational periods of these stars. It is known that the
absolute upper limit on stable stellar rotation is given by the Kepler
frequency, $\Omega_{K}$ , which is the maximum frequency a star can
have before mass loss at the equator sets in \cite{Fridolin1}. Knowing
$\Omega_{K}$ \cite{Fridolin1,Friedman84:a}, the rotational period is
given by $P_K= \Omega_{K}/2 \pi$ .

\section{Results and Conclusions}

\begin{figure}[h]
\includegraphics[width=1.0 \textwidth]{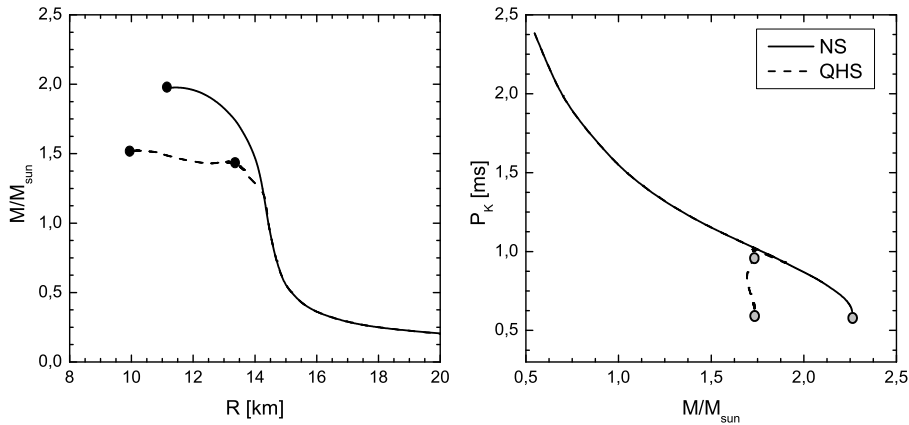}
\caption{Left panel shows the $M-R$ relation for non-rotating NS
  (solid line) and QHS (dashed line). Right panel shows the Keppler
  period $P_K$ as a function of the star mass in solar masses for the
  NS (solid line) and the QHS (dashed line).}\label{MR}
\end{figure}

In this section we show the differences between a standar NS (HV EoS)
which includes neutrons, protons, hyperons, electrons, and muons, and
a magnetized QHS. The left panel of Fig.\ \ref{MR} shows the
M-R relationship for non-rotating stars. Note that for the QHS
there are two mass peaks, one at $1.44 ~M_\odot$ and the other at
$1.52~M_\odot$.  The radii of these stars are $\sim 13$~km and $\sim
10$~km respectively. In contrast to this, the traditional (HV) NS has
a maximum mass of around $\sim 2~ M_\odot$ with an associated radius
of $\simeq 11$~km. We note that not all stars between the two QHS
maxima are gravitationally stable, which is a consequence of the phase
transition between hadronic matter and quark matter
\cite{glendenning,Kettner}. The differences in the masses reflect the
fact that the QHS is more compressed than the NS, since its EoS is
distinctly softer. In the right panel of Fig.\ \ref{MR} we show the
Kepler period as a function of rotating star mass. Rotation shifts the
mass peaks from $\sim 1.44 ~M_\odot$ to $\sim 1.69 M_\odot$ and from
$1.52  ~M_\odot$ to $\sim 1.73 ~M_\odot$ for the QHS, and from $\sim
2 M_\odot$ to $\sim 2.26 M_\odot$ for the NS. Since rotation
stabilizes a star against gravitational collapse, the rotating stars
can carry more mass than the non-rotating star. The range of observed
NS masses is between about 1 and $2~M_\odot$ \cite{zhang} and the
observed rotational periods are greater than 1.38~ms, which is
compatible with the $P_{\rm K}$ curves shown in Fig.\
\ref{MR}. However, phase transition in the cores of NS may lower this
value \cite{Fridolin1}. We obtain $P_{\rm K}=0.88$~ms and $P_{\rm
  K}=0.64$~ms for the stellar QHS twins, and $P_{\rm K}=0.62$~ms for
the maximum-mass NS configuration. 

\begin{figure}[t]
\includegraphics[width=0.9 \textwidth]{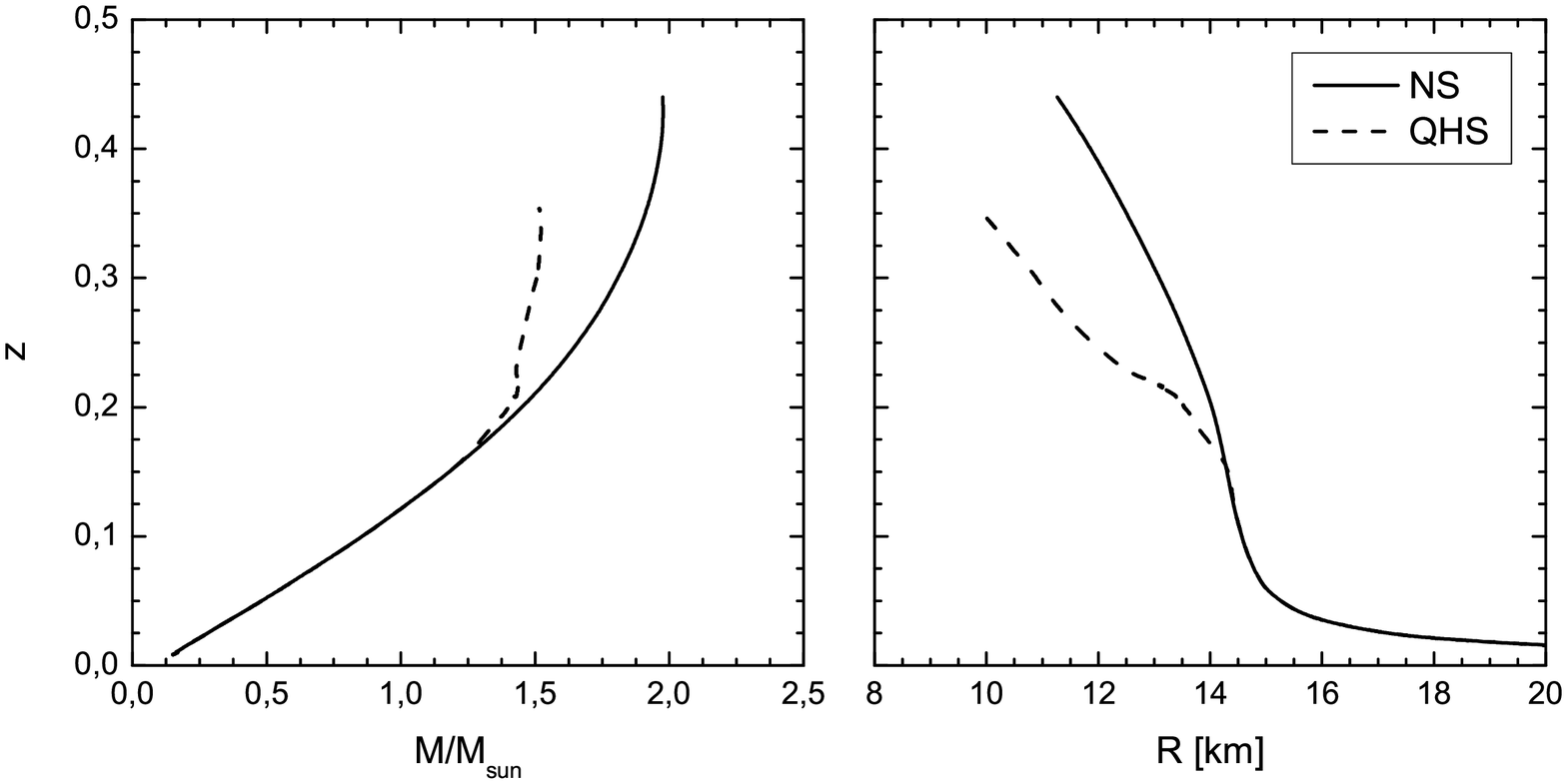}
\caption{Surface redshift parameter z as a function of the star mass in
  solar masses (left panel) and the star radius $R$ (right panel) for
  the NS and the QHS (solid and dashed line
  respectivelly).}\label{zeta}
\end{figure}

In Fig.\ \ref{zeta} we show the gravitational surface redshift. The
differences in the M-R relationships of QHS and NS leads to
markedly different redshift, which opens up the possibility of
distinguishing a QHS from a NS. For the maximum mass QHS we find
$z=0.22$ and $z=0.34$. These values are $\sim 50\%$ and $\sim 20\%$
lower than the redshift of the corresponding NS, which is
$z=0.43$. The inclusion of the magnetic fields, up to $\sim
10^{17}$~G, in the SQM EoS influences the stellar properties discusses
above only very little. This may be different for magnetic fields
greater than $\sim 10^{17}$~G. The study of such high fields is however
outside of the scope of the perturbative formalism presented in this paper.

\section*{Acknowledgements}

M.O. thanks CONICET, the Fulbright Commission, and CIES (Council for
International Exchange of Scholars) for financial support. This work
is supported by the (US) National Science Foundation under Grant
PHY-0854699.

\end{document}